\documentclass[cits]{PoS}
\pdfoutput=1

\usepackage{amsmath}
\newcommand{\FC}{\;,}
\newcommand{\FD}{\;.}
\newcommand{\I}{\mathrm{i}}  
\newcommand{\be}{\begin{equation}}
\newcommand{\ee}{\end{equation}}
\newcommand{\bea}{\begin{eqnarray}}
\newcommand{\eea}{\end{eqnarray}}

\newcommand{\qbar}{\overline{q}}

\title{$DK$ and $D^* K$ scattering near threshold}

\ShortTitle{$DK$ and $D^* K$ scattering near threshold}

\author{\speaker{C. B. Lang}\\
        Institut f\"ur Physik,  University of Graz, A--8010 Graz, Austria\\
        E-mail: \email{christian.lang@uni-graz.at}}
\author{Luka Leskovec\\
        Jozef Stefan Institute, 1000 Ljubljana, Slovenian\\
        E-mail: \email{luka.leskovec@ijs.si}}
\author{Daniel Mohler\\
        Fermi National Accelerator Laboratory, Batavia, Illinois  60510-5011, USA\\
        E-mail: \email{dmohler@fnal.gov}}
\author{Sasa Prelovsek\\
        Department of Physics, University of Ljubljana and Jozef Stefan Institute, 1000 Ljubljana, Slovenia\\
        E-mail: \email{sasa.prelovsek@ijs.si}}
\author{R.~M.~Woloshyn\\
	TRIUMF, 4004 Wesbrook Mall Vancouver, BC V6T 2A3, Canada\\
	E-mail: \email{rwww@triumf.ca}}

\abstract{We study the three $D_s$ quantum channels $J^P = 0^+$, $1^+$ and $2^+$ where
experiments have identified the charm-strange states $D^*_{s0} (2317)$,
$D_{s1}(2460)$, $D_{s1}(2536)$ near the $DK$ and $D^*K$ thresholds, and
$D^*_{s2}(2573)$. We consider correlation functions for sets of $\overline q q$
operators and, for $J^P = 0^+$, $1^+$, also the $DK$ and $D^*K$ meson-meson
interpolators and determine for these cases values of the elastic
scattering amplitude. Constructing the full set of correlators requires
propagators which connect any pair of lattice sites. For one ensemble of
gauge configurations ($32^3\times 64$, $m_\pi\approx 156$ MeV) a
stochastic distillation variant is employed and for another ensemble
($16^3\times 32$, $m_\pi\approx 266$ MeV) we use the full distillation
method. Both, $D^*_{s0} (2317)$ and $D_{s1}(2460)$, are found as bound
states below threshold, whereas $D_{s1}(2536)$, and $D^*_{s2}(2573)$ are
identified as narrow resonances close to the experimental masses.
}

\FullConference{The 32nd International Symposium on Lattice Field Theory,\\
		23-28 June, 2014\\
		Columbia University New York, NY}

\begin{document}

\section{Motivation}
The $DK$ and $D^*K$ scattering process is a technical challenge for Lattice Gauge Theory (LGT) studies as it involves three
quark species $u/d, s$ and $c$ (assuming $u$ and $d$ as degenerate in mass). One has to
consider the light quark mass and at the same time the relatively heavy charm quark.

\begin{figure}[b]
\begin{center}
\includegraphics*[width=0.64\textwidth,clip]{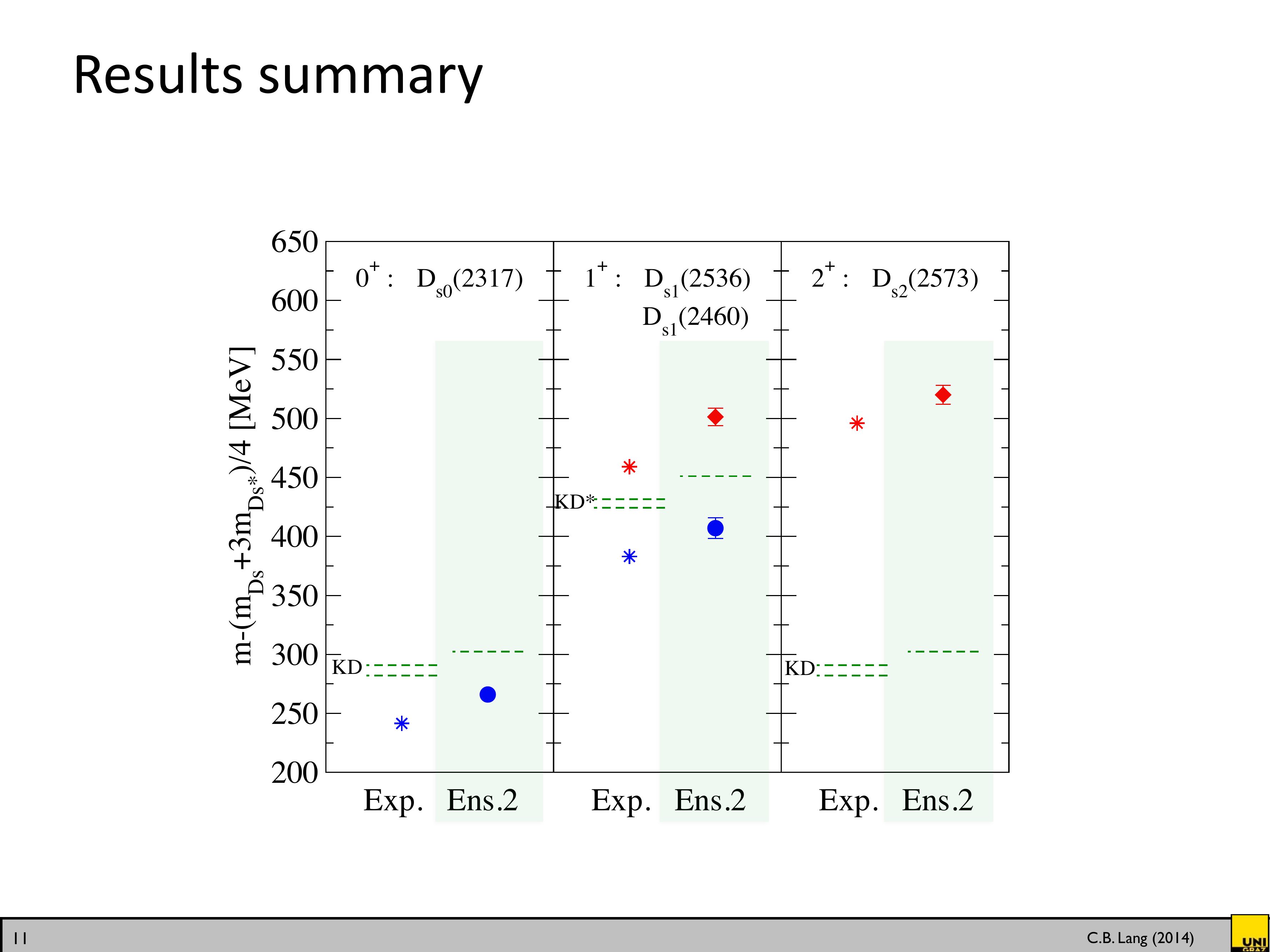}
\end{center}
\caption{Energy differences to the spin-averaged ground state mass. We compare results from experiments (l.h. columns) with our results (r.h. columns).  The dashed green lines denote the relevant $DK$ and $D^*K$  thresholds.}
\label{fig:summary}
\end{figure}

On the other hand it provides an important test on the reliability of such lattice calculations 
near a meson-meson threshold. The quantum channels $J^P=0^+$ and $1^+$ are both
$s$-wave combinations of $DK$ and $D^*K$, respectively. In experiments \cite{Beringer:1900zz} one finds 
states below threshold ($D^*_{s0} (2317)$ and $D_{s1}(2460)$) as well as above 
($D_{s1}(2536)$, and $D^*_{s2}(2573)$ in $2^+$) (see left hand columns in Fig. \ref{fig:summary}). This behaviour was not reproduced in
quark model or in LGT calculations. In both approaches the bound states moved above threshold becoming resonances
\cite{Godfrey:1985xj,Namekawa:2011wt,Mohler:2011ke,Bali:2012ua,Bali:2011dc,Moir:2013ub,%
Kalinowski:2013wsa,Wagner:2013laa}. These calculations were in the ``single hadron'' approach, i.e., without considering the
two-meson channels. As had been pointed out already earlier  \cite{vanBeveren:2003kd} threshold effects may be critical.

Here we report on a study of the coupled $D_s^*$, $DK$ and $D^*K$ systems in the quantum sectors $J_P=0^+$, $1^+$ and
$2^+$. Details can be found in \cite{Mohler:2013rwa,Lang:2014yfa}.

\section{Setup}

For this study two independent ensembles of gauge configurations were used:
\begin{description}
\item[Ensemble  (1)] has $N_f=2$ dynamical light quarks (improved Wilson fermions), a pion mass of 266 MeV, lattice spacing of 0.1239 fm
and lattice size $16^3\times 32$..
 It was produced in a re-weighting study \cite{Hasenfratz:2008fg,Hasenfratz:2008ce}.  
For this ensemble we use the standard distillation method  \cite{Peardon:2009gh} with a complete set of perambulators (one for each time slice set of 96 source vectors). The $s$ and $c$ quarks were both treated as valence quarks only.
\item[Ensemble (2)] with $N_f=2+1$ dynamical quarks has been generated by the PACS-CS collaboration \cite{Aoki:2008sm}. Sea and
valence quarks are non-perturbatively improved Wilson fermions. It has lattice spacing 0.0907 fm, size $32^3\times 64$ and a pion mass of 156 MeV. Here we used the stochastic distillation method \cite{Morningstar:2011ka}. The light and strange quarks were 
dynamical, the charm quark is treated as a valence quark.
\end{description}
The Wick contractions for the $(\bar s \,c)\leftrightarrow DK$ contributions involve partially disconnected graphs with back-tracking quark lines. Both distillation methods proved to be  efficient to compute these. The quark mass parameter for the strange quark was obtained by tuning to the $\phi$ mass and the $\eta_s$ mass.

For the $K$ we use the relativistic dispersion relation and for $D$, $D^*$ the Fermilab method  \cite{ElKhadra:1996mp,Oktay:2008ex} was employed like in \cite{Mohler:2012na}. 

When studying meson-meson scattering in LGT all information is encoded in the discrete energy levels of the
eigenstates of the cross-correlation matrix $C_{ij}(t)=\langle O_i(t)Oâ^\dagger_j(0)\rangle$ between the lattice interpolators
$C_i$. The generalised eigenvalue problem allows the determination of the eigenstates $|n\rangle$ \cite{Michael:1985ne,Luscher:1985dn,Luscher:1990ck,Blossier:2009kd}. From the exponential decay of the eigenvalues one obtains the
lowest energy levels and from the eigenvectors one reconstructs the overlap factors $Z_i^n\equiv\langle O_i|n\rangle$.
The latter provide information on the relative importance of the used lattice operators for the eigenstate.
For details of the fit methods and fit ranges for the energy eigenvalues see 
\cite{Lang:2014yfa}. 

For the $0^+$ channel we used four operators of the type $(\bar s\, A \,c)$ where $A$ denotes a suitable combination of Dirac matrices and
lattice derivatives in a representation of $A_1^+$; there may be mixing with $4^+$ but this is neglected. Furthermore, we
used three operators of type $DK$ with and without relative momenta. Equivalently, for $1^+$ up to eight 
operators of the type $(\bar s\, A \,c)$ and three $D^*K$ interpolators were considered. For $2^+$ we used only two type $(\bar s\, A \,c)$  operators (see App. A of \cite{Lang:2014yfa}).

Partial wave unitarity implies that  the relativistic, elastic scattering amplitude $T(s)$ can  be written as
\be
T^{-1}(s) = \frac{p}{\sqrt{s}} \cot \delta(s) - \I  \frac{p}{\sqrt{s}}\FC
\ee
where $p(s)$ is the momentum and $s=E^2$ the CMS energy squared.\vspace*{9pt}

\noindent\begin{minipage}[lt]{0.45\textwidth}
\quad \quad The figure shows the real part of an inverse elastic partial wave scattering amplitude
$\mathrm{Re}[T^{-1}]$, The straight line is an
effective range approximation to the measured energy level values for $p \cot \delta(p)$ as discussed in the text. Below the threshold one has to add  $|p|/\sqrt{s}$ which is the continuation of the (above threshold)
 imaginary phase space factor $-\I p/\sqrt{s}$ to the upper Riemann sheet below threshold.
 \end{minipage}
\begin{minipage}[rt]{0.5\textwidth}
\quad \includegraphics[width=\textwidth,clip]{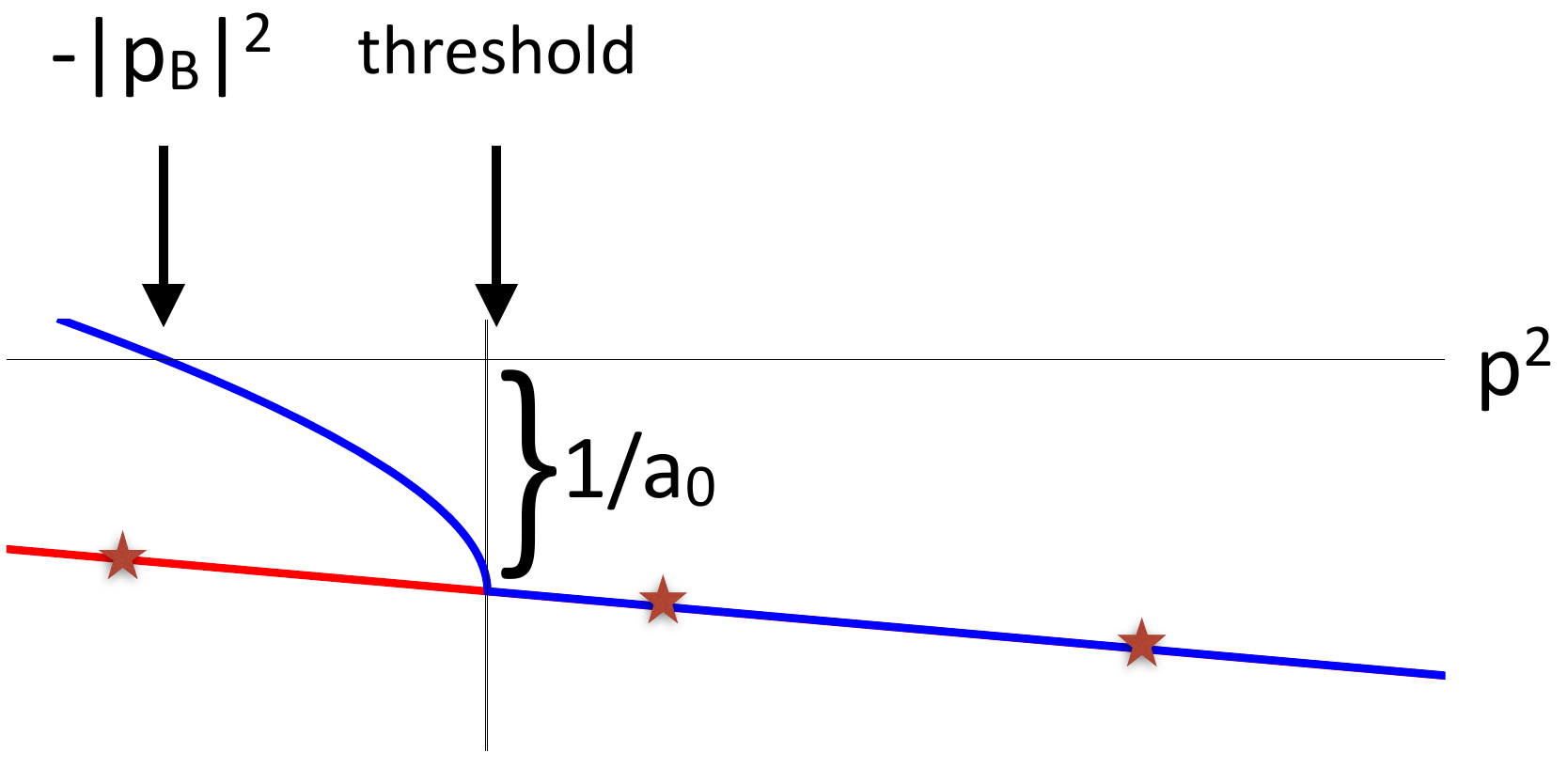}
\end{minipage}\vspace*{9pt}

Near threshold (in the elastic regime) the energy levels $E_n$ in finite volume are related to the values of the real 
part of this inverse scattering amplitude \cite{Luscher:1985dn,Luscher:1986pf,Luscher:1990ux,Luscher:1991cf},
\be
f(p)\equiv p \cot \delta(p)= \frac{2 \mathcal{Z}_{00}(1;(\tfrac{pL}{2\pi})^2)}{L \sqrt\pi}\FD
\ee
This real function has no threshold singularity and the measured values can be found indeed above and below
threshold. For $s$-wave scattering an effective range approximation,
\be
f(p)\approx\frac{1}{a_0}+\frac{1}{2}r_0 p^2\FC
\ee
may be used to interpolate between the closest points near threshold.
The imaginary contribution to $T^{-1}$ becomes real below threshold.
Thus $T^{-1}$ develops a zero where $f(\I|p_B|)+|p_B|=0$. That zero
below threshold corresponds to a bound state pole of $T$ in the upper Riemann sheet, as expected for a molecular
bound state. 

\section{Results}

In Fig. \ref{fig_comp_A1} we compare the energy levels obtained for four subsets of 
correlators. The order of interpolators is listed in App. A of \cite{Lang:2014yfa}.
Set 3 includes interpolators of type $D(0)K(0)$ whereas set 1 has only interpolators
of type $c \bar s$. One finds a phenomenon already observed, e.g., in $\pi N$ scattering in
$J^P=\frac{1}{2}^-$ \cite{Lang:2012db}. When the two hadron interpolators are omitted the energy eigenstate averages 
the two nearby states seen in the complete basis. 

The operator content of the eigenstates is exhibited in the overlap factors $Z_i^n$. These have undefined normalisation. This however cancels in
the ratios $Z_i^n/\max_m Z_i^m$ which give the relative importance of the lattice operators to the eigenstate. 
Fig. \ref{fig_comp_T1} demonstrates the situation: Lattice operator 9 ($D^*(0)K(0)$) is most important for eigenstate 3 whereas
lattice operator 11 ($D^*(1)K(-1)$) dominates eigenstate 4.

  \noindent
  \makebox[\textwidth][t]{ 
         \noindent
         \vbox{\hsize=0.4\textwidth 
            \begin{center}
              \includegraphics[width=0.42\textwidth,clip]{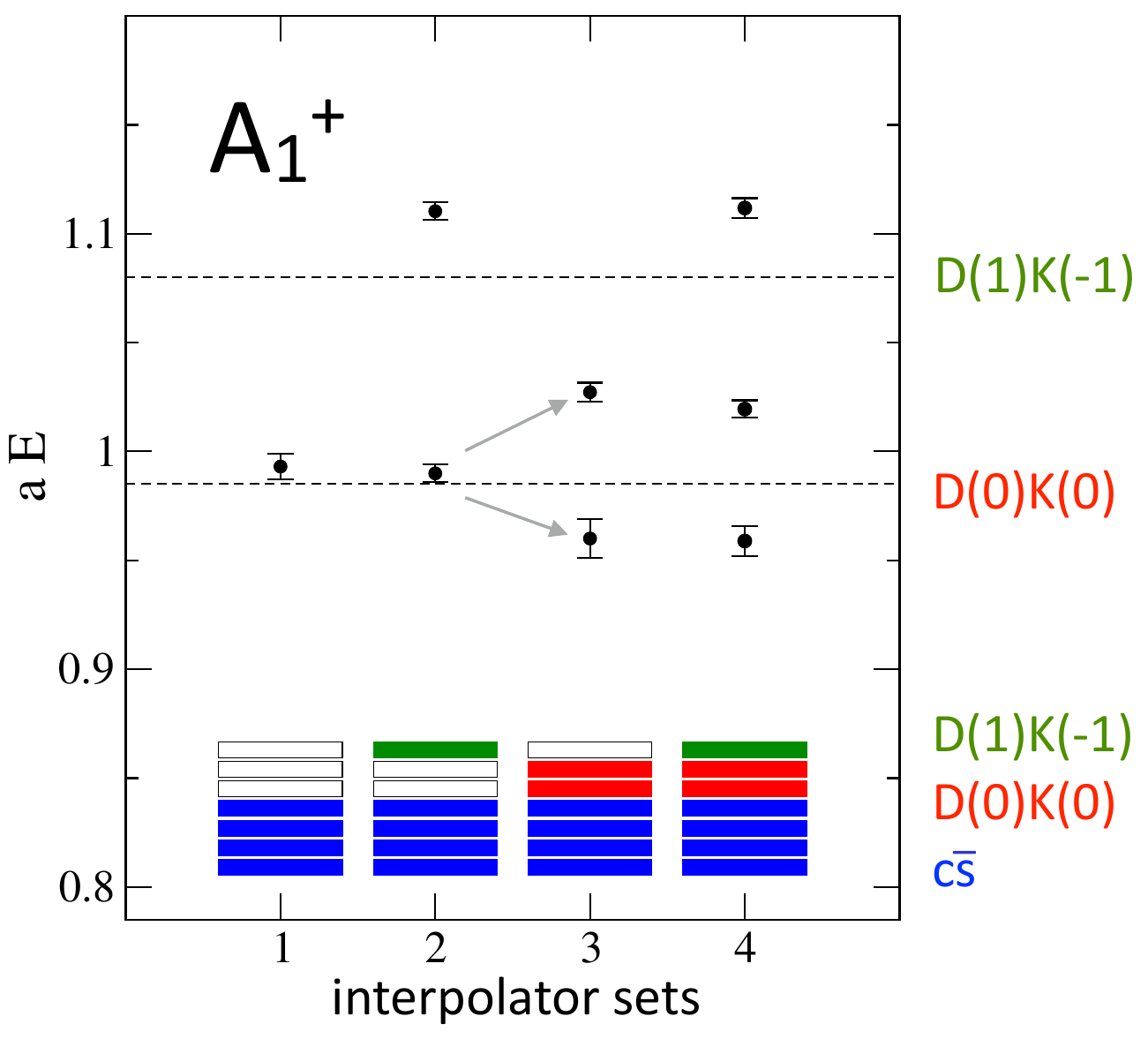}
              \refstepcounter{figure}
              \label{fig_comp_A1}
           \end{center}
            {\sloppy
             {\bf Figure\,\thefigure:\enspace}\raggedright (up)
            Comparison of the energy levels obtained for four subsets of 
             correlators for $J^P=0^+$ (irrep $A_1^+$).}\\
              \refstepcounter{figure}%
              \label{fig_comp_T1}%
                         {\sloppy
             {\bf Figure\,\thefigure:\enspace}\raggedright (r.h.s.)
            Results for energy levels obtained for various interpolator subsets ($J^p=1^+$, irrep $T_1^+$) and ratios of 
overlap factors for one specific case.}
            } 
            \hfill
          \vbox{\hsize=0.6\textwidth 
            \begin{center}
              \includegraphics[width=0.56\textwidth,clip]{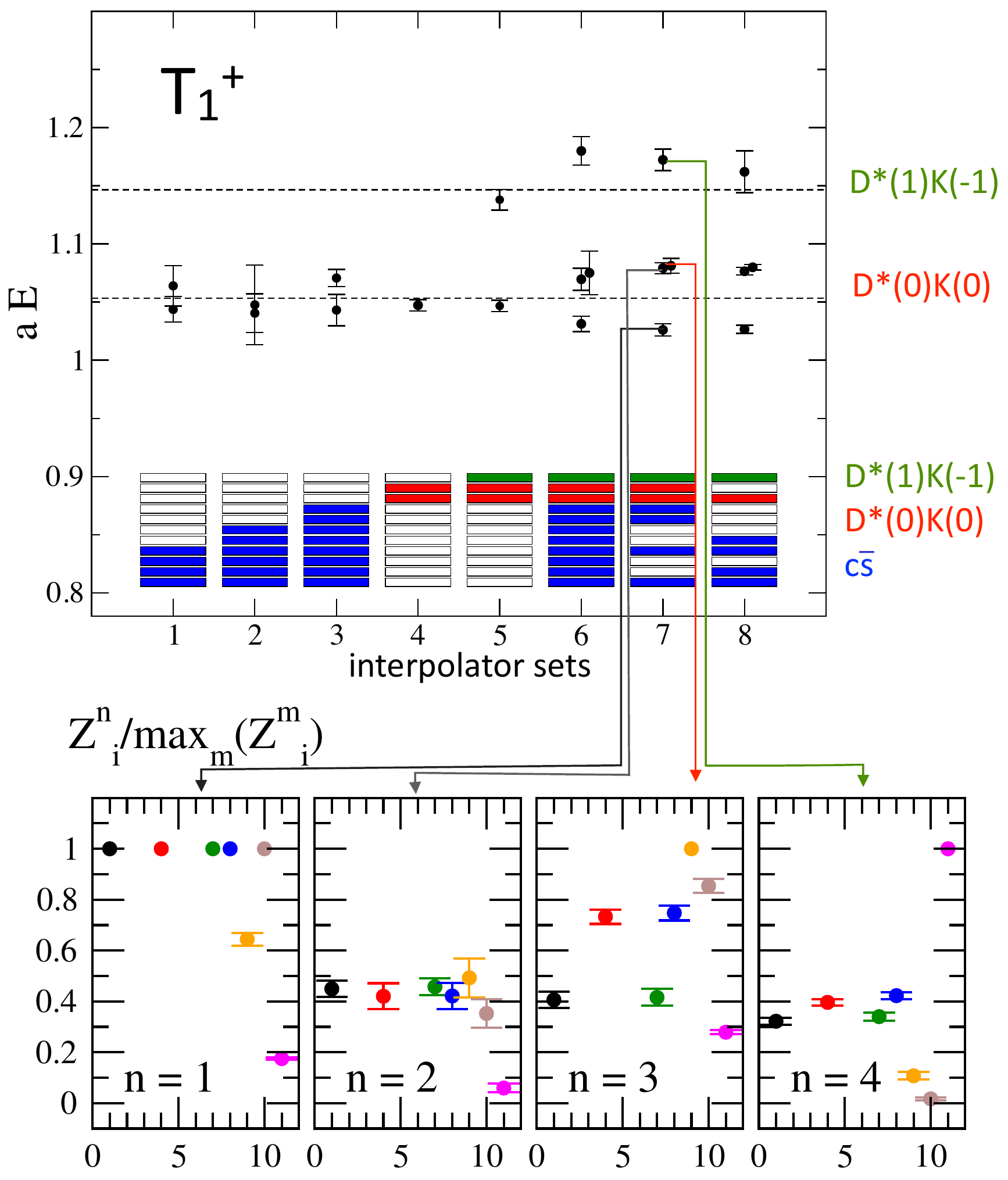}
            \end{center} } 
 }

\begin{table}[tbp]
\begin{center}
\begin{tabular}{l|cc|cccc}
\hline                               
ensemble/irrep & $a_0$ [fm]& $r_0$ [fm] &  $m_K+m_{D}-m_{B} $ [MeV]   &$m_{B}-\tfrac{1}{4}(m_{D_s}+3 m_{D_s^*}) $  [MeV]  \\
 \hline   \hline                              
\multicolumn{3}{l}{irrep $A_1^+$}\\
\hline                                
ens. (1) set 4&-0.756(25) & -0.056(31)  & 78.9(5.4)(0.8) &287(5)(3)  \\
\hline
ens. (2) & -1.33(20)&0.27(17)  & 36.6(16.6)(0.5) & 266(17)(4) \\                         
\hline
 Exp. $D_{s0}^*(2317)$&&                                          &  45.1   &  241.5\\
 \hline  \hline
\multicolumn{3}{l}{irrep $T_1^+$}\\
 \hline 
ens. (1)& -0.665(25) & -0.106(37)  & 93.2(4.7)(1.0) & 404.6(4.5)(4.2) \\              
\hline                                   
ens. (2) set 7& -1.15(19) & 0.13(22)  & 43.2(13.8)(0.6)& 408(13)(5.8) \\                         
ens. (2) set 8& -1.11(11) & 0.10(10) & 44.2(9.9)(0.6) & 407.0(8.8)(5.8) \\                  
\hline                               
 Exp. $D_{s1}(2460)$&&                                          & 44.7   &  383\\                             
\hline                               
\end{tabular}
\end{center}
\caption{\label{tab:A1T1}Scattering length and effective range computed from the linear interpolation near threshold, and parameters for the position of the bound states. The second uncertainty given for values in MeV corresponds to the uncertainty in the lattice scale $a$. The experimental value of $m_K+m_{D}-m_{B}$ is averaged over $D^+K^0$ and $D^0K^+$ (or $D^{*+}K^0$ and $D^{*0}K^+$,
respectively) thresholds. Where given, the set numbers refer to Figs. 2 and 3.}
\end{table}
\begin{table}[tbp]
\begin{center}
\begin{tabular}{l|cc}                               
ensemble& $m_{D_{s1}(2536)}\!-\!m_K\! -\! m_{D^*}$   [MeV] &$m_{D_{s1}(2536)}\!-\!\tfrac{1}{4}(m_{D_s}\!+\!3 m_{D_s^*})$  [MeV]\\
 \hline 
ens. (1) set 4& -53(12)& 444(12) \\              
\hline                                   
ens. (2) set 7& 56(11)& 507(10) \\                         
ens. (2) set 8& 50(8)& 501(8)  \\                  
\hline                               
 Exp.  $D_{s1}(2536)$   &31&459\\                             
\end{tabular}
\end{center}
\caption{\label{tab_T1_Ds2536}Comparison of the mass of  $D_{s1}(2536)$ with experiment.}
\end{table}

Our main results are values for the bound state  positions and for the scattering lengths.
Table~\ref{tab:A1T1} summarises the bound state results for both ensembles.

The second lowest state in both ensembles is identified with $D_{s1}(2536)$. In ensemble (1) with the heavier pion the state lies below $m_{D^*}+m_K$, but in the ensemble (2) we find it above this threshold. The mass is compared with experiment in Table \ref{tab_T1_Ds2536}.
In the heavy quark limit \cite{Isgur:1991wq} $D_{s1}(2536)$ does not couple to $D^*K$ in $s$-wave. We find that the composition of the states with regard to the $\qbar q$ operators is fairly independent of whether the $D^*K$ operators are included or not. The level is not observed if only $D^*K$ interpolators are used. 

Since the mass of the  $D_{s2}^*(2573)\to DK$ (with a width of 17(4) MeV) is quite far away from the first (in $d$-wave) relevant level $D(1)K(-1)$ we did not include the $DK$ or possible $D^*K^*$ interpolators. 
We find differences $(E-\tfrac{1}{4}(m_{D_s}\!+\!3 m_{D_s^*}))$ of 473(10)(5) MeV (ensemble (1)) and 520(8)(7)  MeV (ensemble (2)), comparable to the experimental value 496 MeV.

\textit{Acknowledgments:}
We thank Anna Hasenfratz and the PACS-CS collaboration for providing gauge configurations. D.~M. would like to thank E.~Eichten, F.-K. Guo, M.~Hansen, A.~Kronfeld, Y.~Liu and J.~Simone for insightful discussions. The calculations were performed on computing clusters at TRIUMF, the University of Graz, NAWI Graz, and at Jozef Stefan Institute. This work is supported in part by the Austrian Science Fund (FWF):[I1313-N27], by the Slovenian Research Agency ARRS project N1-0020 and by the Natural Sciences and Engineering Research Council of Canada. Fermilab is operated by Fermi Research Alliance, LLC under Contract No. De-AC02-07CH11359 with the United States Department of Energy.

\providecommand{\href}[2]{#2}\begingroup\raggedright\endgroup


\end{document}